\documentclass[lettersize,journal]{IEEEtran}
\usepackage{multirow}

\ifCLASSOPTIONcompsoc
  \usepackage[nocompress]{cite}
\else
  \usepackage{cite}
\fi
\usepackage{amsmath,amssymb,amsfonts}
\usepackage{algorithmic}
\usepackage{graphicx}
\usepackage{textcomp}
\usepackage{xcolor}
\def\BibTeX{{\rm B\kern-.05em{\sc i\kern-.025em b}\kern-.08em
    T\kern-.1667em\lower.7ex\hbox{E}\kern-.125emX}}

\usepackage{amsmath}
\usepackage{cite}
\usepackage{graphicx,amssymb}
\usepackage{amsmath,amsfonts,amssymb}

\usepackage{url}
\usepackage{graphicx}
\usepackage{enumitem}

\ifCLASSINFOpdf
\else
\fi
\usepackage{ragged2e}

\begin{document}
%
\title{Quantum Game Theory meets Quantum Networks}
%
%
%
%

\author{Indrakshi~Dey,~\IEEEmembership{Senior Member,~IEEE,}~Nicola~Marchetti,~\IEEEmembership{Senior Member,~IEEE},~Marcello~Caleffi,~\IEEEmembership{Senior Member,~IEEE},~and~Angela~Sara~Cacciapuoti,~\IEEEmembership{Senior Member,~IEEE}
\IEEEcompsocitemizethanks{I.~Dey is with the Walton Institute for Information and Communications Systems Science, South East Technological University, Ireland. (Email: indrakshi.dey@waltoninstitute.ie)\\
N. Marchetti is with CONNECT, School of Engineering, Trinity College Dublin, Ireland. (E-mail: nicola.marchetti@tcd.ie)\\
M.~Caleffi and A.~S.~Cacciapuoti are with University of Naples Federico II, Naples, Italy. (Email: marcello.caleffi@unina.it, angelasara.cacciapuoti@unina.it)\\
The contribution from I. Dey and N. Marchetti is supported by HORIZON-MSCA-2022-SE-01-01 project COALESCE under Grant Number 101130739, HORIZON-EOSC-03-2020 project QCloud under Grant Number INFRAEOSC-03-2020 and ERDF project SAtComm under Grant Number EAPA\_0019/2022.
The work of M. Caleffi was partially supported by PNRR MUR project PE00000001 ``RESTART", while the work of A.S. Cacciapuoti was partially supported by PNRR MUR project PE00000023 ``NQSTI".}
}
\maketitle
\begin{abstract}
Classical game theory is a powerful tool focusing on optimized resource distribution, allocation and sharing in classical wired and wireless networks. As quantum networks are emerging as a means of providing true connectivity between quantum computers, it is imperative and crucial to exploit game theory for addressing challenges like entanglement distribution and access, routing, topology extraction and inference for quantum networks. Quantum networks provide the promising opportunity of employing quantum games owing to their inherent capability of generating and sharing quantum states. Besides, quantum games offer enhanced payoffs and winning probabilities, new strategies and equilibria, which are unimaginable in classical games. Employing quantum game theory to solve fundamental challenges in quantum networks opens a new fundamental research direction necessitating inter-disciplinary efforts. In this article, we introduce a novel game-theoretical framework for exploiting quantum strategies to solve, as archetypal example, one of the key functionality of a quantum network, namely, the entanglement distribution. We compare the quantum strategies with classical ones by showing the quantum advantages in terms of link fidelity improvement and latency decrease in communication. In future, we will generalize our game framework to optimize entanglement distribution and access over any quantum network topology. We will also explore how quantum games can be leveraged to address other challenges like routing, optimization of quantum operations and topology design.
\end{abstract}




%

\section{Introduction} \label{sec:introduction}

%
%
%
%
\IEEEPARstart{E}{veryone} wants to enter the quantum race, from tech giants to states and governments with massive public funds for infrastructure development, like European Commission’s Quantum Technologies Flagship program, and USA’s National Quantum Initiative worth 1.2 billion US. Such a huge investment is motivated by the promise of quantum computer’s capability of executing tasks that choke classical computers within realistic time-scale \cite{3}. 

Unleashing the full potential of quantum computing requires implementation of operations among a large number of qubits, which is impossible for a single quantum processor to execute with the current level of technology. In order to circumvent the challenges associated with large monolithic quantum processors, the most promising approach is to network multiple realizable smaller quantum processors (or nodes) together \cite{4}. Each such processor can execute few operations individually, but when interconnected in a quantum network i.e. the \emph{Quantum Internet}, one is able to compile large and complex computing tasks exceeding the computing power of a single quantum processor. As quantum networks will be rolled out, providing true quantum connectivity between quantum computers over short and long distances, it will be possible to realize a wide range of distributed quantum computing, secure communications, and quantum-enhanced applications.


Quantum networks have to exploit the unique phenomenon of \textbf{\emph{entanglement}}\cite{7} to fully unleash the communication and computing potentialities allowed by quantum mechanics. Entanglement, unique to quantum systems and unmatched in classical physics, is key for quantum communication across distant nodes. It's as vital for quantum networks as frequency bands are for classical networks. Both bipartite and multipartite, entanglement is a fragile, challenging-to-maintain resource. Efficiently managing and distributing it among network nodes is crucial for leveraging its properties in quantum communication, presenting a complex yet fundamental challenge \cite{4}. A promising solution for entanglement distribution within quantum networks can be the development of centralized or distributed decision-making targeting the optimization of long-term system properties. 

Decision-making in networks balances resource distribution and long-term outcomes, optimizing key metrics within specific constraints. In classical networks, constraints include environmental factors like fading and interference, with metrics like error rates and spectral efficiency. In quantum networks, constraints stem from interactions between quantum states and the environment, leading to decoherence—a phenomenon unique to quantum systems. Here, the primary metrics are fidelity and communication rate, measured in e-bits per channel use. \cite{9}.

\subsection*{Motivation}

Classical game theory has proved to be instrumental in optimized decision-making for resource distribution, allocation and sharing within resource-constrained classical networks, like Internet-of-Things (IoT), network of unmanned aerial vehicles (UAVs) \cite{10}. Game theory is preferred for online decision-making in scenarios with limited information, outperforming traditional numerical optimization and learning techniques. It handles large networks and numerous parameters more efficiently than numerical optimization and doesn't rely on extensive pre-existing data like learning methods. Game theory enables adaptable modeling of uncertainties and learning from network topology, facilitating stable, decentralized coordination. Its distributed approach also scales well with network size, offering manageable computational complexity and memory requirements.

\textbf{Quantum Games promise increase in efficiency and payoffs, emergence of new equilibria and novel game strategies which are simply not possible in the classical domain}. Quantum games, leveraging their strategic and rule-based nature, have been utilized to reinterpret various quantum algorithms and information processing techniques, providing deeper insights into these areas. This includes applications such as the quantum version of the Prisoner's Dilemma demonstrated on nuclear magnetic resonance quantum computers, exploring the one-way model of quantum computation, and representing quantum non-locality, cluster-state generation, and various paradoxes through non-zero-sum \cite{11} or graphic games. However, the potential of both classical and quantum games has not yet been fully tapped for addressing specific challenges like the distribution and sharing of delicate resources like entanglement, optimizing network topology, and ensuring high-fidelity information routing in quantum networks, whether in competitive or cooperative settings. \textbf{As the Quantum Internet gradually becomes a reality, it will be possible for quantum networks to leverage the benefits offered by quantum games over classical games in the aforementioned challenges. Indeed, by incorporating quantumness in form of pre-shared entanglement among network nodes, quantum games can achieve equilibria outperforming their classical counterparts and allow the players to explore correlated outcomes (with no-counterpart in the classical world) even in the absence of communication \cite{11a}.} 

\subsection*{Contribution}

In this article, we aim at exploiting the promise of game theory for quantum networks. As a first-ever application, we propose a novel game-theoretic framework for entanglement distribution, capable of establishing stable links between any two nodes separated by a distance within fixed network topologies. Consequently, we investigate how classical and quantum strategies can be formulated for the game framework such that fidelity is maximized, while maintaining entanglement rate, and link latency is minimized subjected to coherence time constraint. 

{In the landscape of quantum games, \cite{11a} initiated the exploration of quantum information processing principles within game theory, laying the groundwork for subsequent advancements. \cite{11} further developed the theoretical understanding of quantum games, becoming fundamental to the intersection of quantum mechanics and game theory. Envisioning a quantum internet, as presented by \cite{7}, added complexity and scalability to applying quantum concepts in distributed systems. The dynamics of quantum games were explored by \cite{7a}, delving into cooperative and competitive aspects in a distributed setting. \cite{7b} extended the discussion to practical applications, emphasizing the role of quantum technologies in game theory. In our seminal work, we harness the transformative potential of quantum games to revolutionize communication and information processing paradigms. Serving as a testing ground, quantum games explore cooperative and competitive dynamics in networked environments, contributing to the development of tailored quantum algorithms for distributed systems. As quantum networks seek efficient information transfer through entanglement, quantum games elucidate strategic aspects, advancing our understanding of optimal quantum resource utilization in networked environments, with implications for the future of quantum communication and computing.}

We formulate two different kinds of games; i) multiplayer coalition game where multiple nodes within a quantum network cooperate to establish entanglement (link) between source and final destination, ii) 2-way consensus game, where each node decides on the next 1-hop destination among multiple nodes available to communicate with. We devise both classical and quantum strategies for each game; where quantum strategies offer advantage in performance in both cases. 

Introduction of quantum strategies blurs the boundary between cooperative and competitive scenario, as the initial entangled quantum state allows players to utilize the correlations present in the state. In this paper, we deviate a bit from this condition in our 2-way consensus game, where the players' action does not depend only on the player's observation of the quantum state received from the referee. The player decides on the next node for communicating, based on fidelity payoff and latency cost estimates over the forwards links available.

\begin{figure}[t]
\includegraphics[width=0.9\columnwidth]{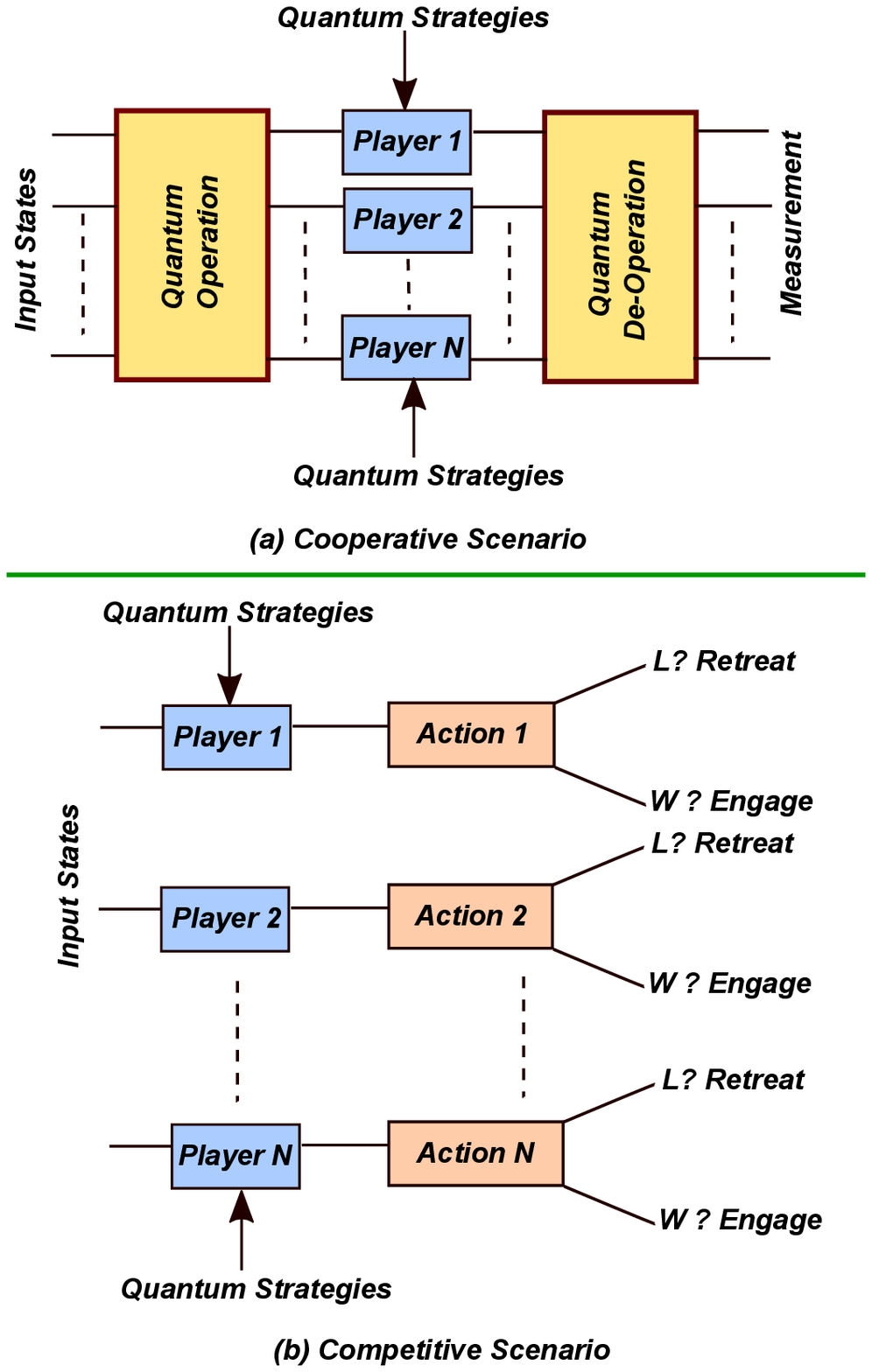}
\caption{Diagrammatic Representation of Quantum Games in Cooperative and Competitive Scenarios. {Here `W' represents winning a particular game-round, while `L' represents losing that particular game-round. Also it is noteworthy that in a cooperative scenario, though individual players employ individual strategies, action is taken jointly by all the players. While in a competitive scenario, individually different actions are taken by individual players.}}
\label{FIG1}
\vspace{-5mm}
\end{figure} 

\section*{Background on Quantum and Classical Games}

Game theory provides a set of mathematical tools and frameworks that leverage interactions of rational heterogeneous self-interested entities with different information availability, to achieve a global objective and predict system-level emerging behaviors. From a network perspective, game models can capture the influence of the network topology on distributed decision-making processes of entities, with the freedom to plan their moves independently based on their own goals and incomplete local information. A basic game model involves five components : a) \textbf{Players} - Participants or decision-makers; b) \textbf{Action} - Preferred move of each player through implementation of player's control; c) \textbf{Information} - Local or global knowledge that dictates each player's decision; d) \textbf{Strategy} - Mapping player's moves with the information available to it at any point in time; e) \textbf{Utility or Payoff} - Each player's preference ordering of possible game outcomes. 

A very important concept in game theory is \textbf{Equilibrium}. The most commonly known form of equilibrium is the Nash Equilibrium (NE). NE represents a set of optimal strategies in a game where no player can improve their expected payoff by unilaterally changing their strategy, assuming complete knowledge of opponents' strategies. In scenarios with incomplete information, this extends to Bayesian equilibrium, using probabilities of various strategy combinations. In network contexts, like forming a link, Wardrop equilibrium \cite{12} ensures minimum information transit time. The choice of equilibrium, influenced by the game type and player nature, is key to determining the best outcome for each player.

Games can be classical or quantum depending on whether they employ classical or quantum resources/strategies respectively. \textbf{Quantum games offer advantages over classical games in terms of winning probabilities, efficiency, payoffs and equilibria} \cite{11a}. Quantum strategies also offer strictly higher average payoffs over classical strategies in competitive scenarios where participating entities have conflicting interests. For example, in CHSH games \cite{13}, if the \emph{a priori} shared resource between two spatially-separated players is classical, the probability of winning is 0.75, while if the resource is quantum (like a pair of maximally-entangles qubits), the probability of joint winning exceeds 0.75 i.e., $\cos^2\pi/8 > 0.75$. This gain in payoffs can be attributed to the fact that entanglement interferes with the \textbf{dilemma} present in classical games. Classical games often present dilemmas where one player's win necessitates another's loss. However, quantum games, introducing entanglement at the outset, allow multiple players to attain satisfactory payoffs, broadening the strategy spectrum beyond classical confines. Quantum strategies, crafted from convex linear combinations of unitary actions, enable several players to simultaneously maximize their payoffs, conforming to Glicksberg's extended version of NE in the quantum realm. \cite{14}. 

\textbf{Quantum games can be cooperative} (all players have common interests) \textbf{or competitive} (players compete for a particular target or have conflicting interests). Cooperative and competitive quantum games differ in player strategies and actions. Cooperative games involve players coordinating strategies through quantum entanglement, enhancing winning probabilities by sharing information on others' moves. In competitive games, players independently decide strategies based on personal circumstances, affecting their outcomes as win, lose, or retreat. These distinct approaches are illustrated in Fig.~\ref{FIG1}, highlighting the diverse dynamics and outcomes of quantum game strategies.

\section*{Game-based Optimization Framework for Entanglement Distribution}

This section focuses on using a game-theoretic approach to address key challenges in quantum networks, particularly in distributing entanglement among nodes to optimize system properties like fidelity, coherence, entanglement rate, and communication latency. Given current technology limitations, quantum networks often have fixed topologies with optical fiber-linked nodes. These nodes form coalitions for efficient computing tasks, requiring entanglement distribution across them. The distribution process is influenced by varying coherence times of links and aims to minimize latency and maintain fidelity and entanglement rate within the network's coherence time. This approach is crucial for optimized network performance, regardless of the network's diverse applications.
\begin{figure}[t]
\includegraphics[width=0.99\columnwidth]{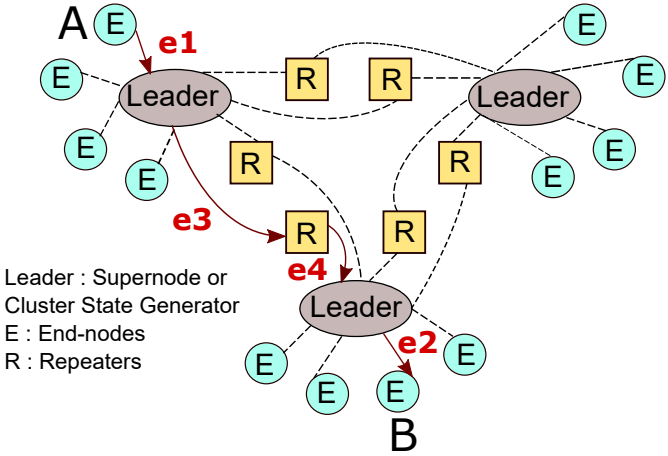}
\vspace{-3mm}
\caption{Optimized information and resource flow over a quantum network topology with three leader nodes (Leader), multiple repeaters (R) and end-nodes (E) between A and B; both classical and quantum coalition games are employed and $e_1 \to e_3 \to e_4 \to e_2$ are selected links for information flow.}
\label{FIG2}
\end{figure}

Our framework utilizes a game-theory approach to optimize entanglement distribution in a fixed quantum network topology. Here, quantum nodes act as players, and their utility is the difference between the entanglement rate and fidelity (the payoff) and the latency of the link (the cost), constrained by coherence time. Nodes assess local fidelity and entanglement rates, adjusting strategies to minimize latency within the coherence time, ensuring an equilibrium flow of entanglement across the network. This model exclusively involves quantum players, reflecting the quantum nature of the network. We, however, explore and compare classical and quantum strategies and resort to \emph{Nash equilibrium} for the classical version and to \emph{Wardrop equilibrium} for the quantized version. {For our game framework, we define 
\begin{itemize}[leftmargin=*]
\item \textbf{Utility Function -} $U_i = \text{Payoff}_i~-~\text{Cost~Function}_i$ 
\item \textbf{Payoff and Cost Function Components -} $\text{Payoff}_i = \text{Fidelity}_i + \text{Entanglement Rate}_i$ \\
$\text{Cost~Function}_i = \text{Link~Latency}_i$
\item \textbf{Quantum Strategy Choice -} \\
$\text{Strategy~Choice of $i$th node} = \text{Unitary~Rotation~on~Qubit}$
\end{itemize}}

{\textbf{\emph{Assumptions}} in our model involve a fixed quantum network topology, where nodes act as players forming coalitions for efficient task execution, and entanglement is distributed among coalition nodes. Coherence time heterogeneity among links introduces variability in the entanglement distribution process, with a primary focus on optimizing long-term system properties, including fidelity, coherence, entanglement rate, and communication latency. We also assume manipulation of entangled qubits through arbitrary unitary rotations for quantum nodes strategies, influencing coalition participation and entanglement distribution. However, this means that our framework is limited to fixed topologies. This is reasonable given the early stage of quantum network technology. \textbf{\emph{Limitations}} highlight the study's tailored scope to fixed-topology quantum networks, the idealized nature of quantum strategies focusing on entangled qubits, and the abstract nature of equilibrium notions such as Nash and Wardrop equilibria, which, while representing stable solutions within our model, may face practical constraints in real-world implementations.}


Another important aspect for practical implementation is to search for a stable solution. Wardrop equilibrium is analogous to NE, however, we consider it for the quantum strategies, in which case, the nodes aim at equalizing latency over their individual forward (outgoing) links. It is worth-noting, we consider the outgoing links from each node in order to account for the constraint on the link coherence time. Quantum strategies start with each node (player) being allocated a single entangled qubit. The arbitrary unitary rotation that nodes apply to their qubit is their strategy choice. The strategy choice determines whether a particular node will be part of the coalition to which entanglement will be distributed.  

In quantum games, Wardrop equilibrium, traditionally linked with classical traffic, is adopted to optimize entanglement distribution strategies, considering coherence time and quantum mechanics complexities. While classical Wardrop equilibrium minimizes travel time for equalized routes; in quantum networks, players distribute entanglement, factoring in distribution route dynamics and strategic behavior. Computing Nash equilibrium in quantum games is challenging due to entanglement and no-cloning properties, while Wardrop Equilibrium is easier to compute and implies stable entanglement distribution, shedding light on quantum strategy stability. It is worth-mentioning here that the validity of conclusions hinges on choosing the appropriate equilibrium concept that aligns with the dynamics of the specific problem under analysis. Computing Wardrop equilibrium in quantum network problems, like entanglement distribution, will accurately reflect individual nodes' behavior in optimizing entanglement distribution routes based on minimization of travel time.

\subsection*{Scenario 1}

We consider a network topology which consists of $N$ super-nodes (leader nodes), each capable of generating a given $M$-partite cluster state. Each super-node is connected to $M$ end-nodes. There are also $L$ repeater nodes between each pair of super-nodes. We represent such an example topology in Fig.~\ref{FIG2} with $N = 3$, $M = 4$ and $L = 2$. Let us assume we want to establish a communication path between source A and destination B. Consequently, we want to establish the best possible link between A and B to distribute entanglement in a way that i) minimizes the number of quantum operations and latency in entanglement distribution, ii) maximizes fidelity within the coherence time of the link and iii) maintains the target overall network entanglement rate.

In our scenario, we model coalition formation as a game, where links are based on the coherence time between source and destination. Shorter coherence times require distributing entanglement over fewer hops. We use entanglement rate as the payoff and number of hops as the cost, to optimize the coalition of nodes for link setup. This is achieved through iterative coalition formation and entanglement distribution until a stable coalition is formed, after which the established link is used for various tasks.
\begin{figure}[t]
\includegraphics[width=0.9\columnwidth]{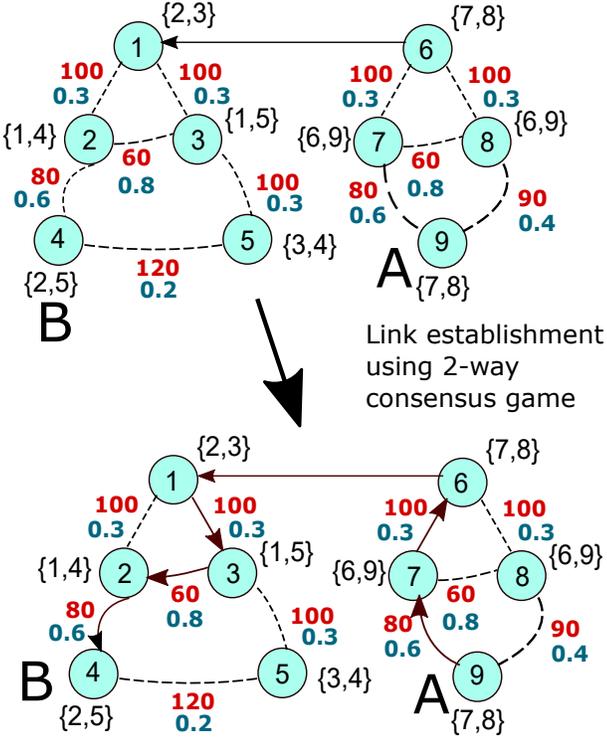}
\caption{Optimized information and resource flow through the control of the next one-hop link within a tree-like quantum network topology; nodes 1 and 6 are the leader nodes that are connected through a fixed link -- Links are selected for information exchange from $A$ to $B$ using a 2-way consensus game. Numbers in red {(e.g., $60, 80, 90, \dotso$)} are latency-based cost and numbers in {teal (e.g., $0.3, 0.4, 0.5, \dotso$)} are fidelity-based payoffs. {The numbers inside $\{...\}$ represent the identities of the possible next hops, either of which the present node can connect to in the next step.}}
\vspace{-3mm}
\label{FIG3}
\end{figure}

 \subsection*{Scenario 2}

Here, let us consider a tree-like network topology which consists of multiple trees. Each tree consists of one source node and several leaf nodes. However, each source (leader) can exchange information with only one destination leaf node at any point of time. We represent such an example topology in Fig.~\ref{FIG3} with 2 trees, where one tree has 5 leaf nodes and other one consists of 4 leaf nodes. In this case, we want to establish the best possible path for entanglement distribution between two leaf nodes A and B, where A and B belong to two different trees, in a way that, i) latency in communication is minimized, ii) the link fidelity is maximized and iii) the entanglement distribution can be completed within the coherence time of the link.

In order to optimize the overall network performance, each quantum node needs to decide on the next 1-hop destination to communicate towards, depending on its current state-related information (location, direction). All such 1-hop links between the source and the final destination will form the link topology. We consider a two-way choice for each node; an example is provided in Fig.~\ref{FIG3}. Let node 2 decides to switch its link from node 1 to node 3. The link between nodes 1 and 2 is removed followed by a consensus between nodes 2 and 3 to establish the link. Through this link deviation, the latency cost is reduced from 100 to 60 and the fidelity payoff increases from 0.3 to 0.8. So the edge between two nodes in this case dictates the two-way consensus game process.

\subsection*{Classical V/s Quantum Strategies}

Here we describe the differences between classical and quantum strategies for each of the two scenarios. For scenario 1, we apply both the classical and quantum forms of the multiplayer coalitional game, towards solving the optimized link formation between two quantum end-nodes in the network topology outlined in Fig.~\ref{FIG2}. In the classical form of the game, based on the classical strategy adopted by each player, one guarantees that each player forming the coalition is rewarded by a certain amount called the `value of coalition'. Other players in the game who are unable to join the coalition can prevent the players forming the coalition from getting any more payoffs than the `value of coalition'. In our particular network topology set-up, the `value of coalition' is attributed to a target network throughput. For the quantum version of the game, the leader node to which the source node $A$ is connected, is selected as the referee or arbiter of the game. The referee prepares an initial quantum cluster state and forwards it to the players. Each player is in possession of a single entangled qubit, on which it employs an arbitrary unitary rotation depending on its preferred action. The resultant state is forwarded back to the referee for measurement and the corresponding payoff assignment. If the initial quantum states are unentangled, the quantum coalition game breaks down into its classical form.

For scenario 2, we apply both classical and quantum versions of the multiplayer 2-way consensus game for optimizing 1-hop link control between nodes within the network topology outlined in Fig.~\ref{FIG3}. In the classical game, players chooses the next hop from two options, seeking to minimize costs and maximize payoffs in subsequent moves. While individual choices are autonomous, other nodes in the network can affect the overall utility but not specific player decisions. The quantum version involves a fair coin flipping game, ensuring no cheating. Players are aware of and agree on each other's decisions regarding link formation and game outcomes. This setup allows each player to track their progress and ensures convergence of the game, even with multiple players independently deciding their moves. {It is worth-mentioning here that, we can also consider relaxation of the `no-cheating' \cite{17} requirement; an essential generalization that we are actively exploring and and intend to incorporate into our upcoming work.}
\begin{figure}[t]
\includegraphics[width=0.99\columnwidth]{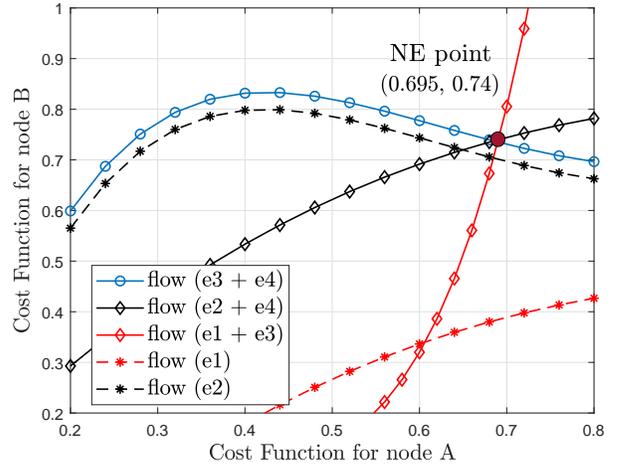}
\caption{Nash equilibrium point between the links for information flow from $A$ to $B$ with the objective of minimizing the number of quantum operations and latency, such that the information is exchanged within the end-to-end coherence time of the link. The cost function for node $A$ and node $B$ is computed using the total latency experienced over all the links that information of the nodes propagates on. These results are based on topology outlined in Fig.~\ref{FIG2}.}
\label{FIG2a}
\end{figure} 

\section*{Results and Analysis}

To implement the proposed game-based optimized link set-up for information flow and resource access within quantum networks, we conduct numerical evaluations. For each set of parameter settings, simulations are run through 1000 trials and the results are averaged out. {Since the average lifetime of a qubit with current superconducting technology is around 500$\mu$s \cite{16}, we employ a synchronization time step of 300$\mu$s}. The probability that a link will exist between any two quantum nodes, irrespective of their type, repeater, end-node or leader, is expressed as $p(m,n) = \mu \exp[-d(m,n)/\delta \lambda]$, where $d(m,n)$ is the Euclidean distance between nodes $m$ and $n$, $\delta$ is the maximum distance between $m$ and $n$, $\mu$ and $\lambda$ are the control parameters of the distribution; $\mu, \lambda \in (0,1]$; $\mu$ controls the number of edges (links) present in the network topology and $\lambda$ controls the length of different links.
\begin{figure}[t]
\includegraphics[width=0.99\columnwidth]{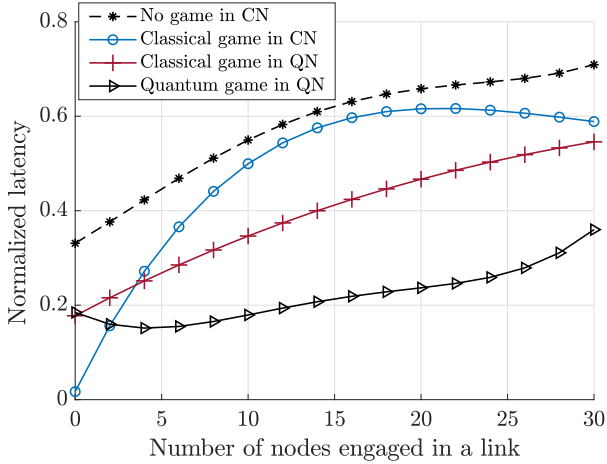}
\vspace{-4mm}
\caption{Variation in the normalized delay experienced by information flow arriving at any end-node within the topology obtained in Fig.~\ref{FIG2}, as a function of the increasing number of nodes simultaneously communicating over the network. CN stands for Classical Networks and QN stands for Quantum Networks.}
\label{FIG2b}
\end{figure}

In Fig.~\ref{FIG2a}, we analyze the Nash equilibrium for the latency and operations minimization problems with $X$ and $Y$ axes representing the cost incurred at nodes $A$ and $B$ respectively. The curves are the best response functions or the information exchange rate over the edges. We are particularly interested in the equilibrium point over the edges between the leader-repeater-leader nodes. We reached a unique Nash equilibrium point at (0.695,0.74).
\begin{figure}[t]
\includegraphics[width=0.99\columnwidth]{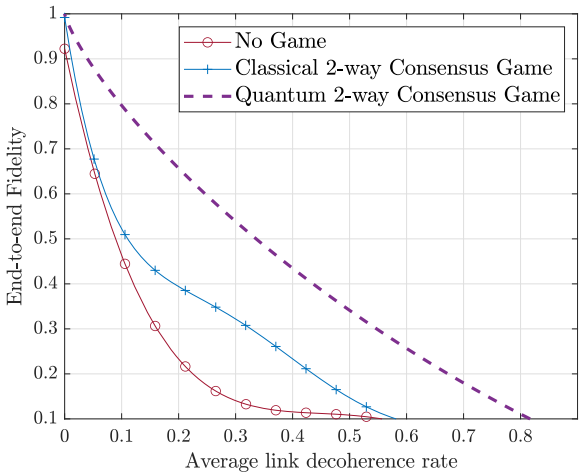}
\vspace{-4mm}
\caption{End-to-end fidelity improvement with variation in the link decoherence rate through the application of classical 2-way consensus game and its quantized version. These results are based on the topology outlined in Fig.~\ref{FIG3}.}
\label{FIG4}
\end{figure} 

Next we investigate the efficiency of the classical and quantum versions of the coalition game approach for topology extraction in quantum networks in Fig.~\ref{FIG2b}. We represent the normalized latency in one communications cycle as a function of the number of nodes present in the network. The average latency is calculated by summing the delay experienced over all the links and then dividing the summation by the number of hops. Performance of four different scenarios is compared, where no game is used for classical networks, classical game for classical networks, classical game for quantum networks, and quantum game for quantum networks. The quantum strategies emerge as the winner and the reason can be explained through an intuitive example. {It is worth-noting here that in quantum networks, reduced latency ensures sustained entanglement amidst environmental challenges. For quantum operations, lower latency permits rapid gate operations and reduces errors. Real-time quantum processing benefits from quicker decision-making, and lower latency supports scalability, crucial as quantum technologies advance and systems grow in complexity. Minimizing latency is therefore paramount for maintaining coherence and reliability across expanding quantum systems.}

Let us consider a coin toss scenario. Classical players choose heads or tails with equal chances. In the quantum version, entangled photons replace coins, and players use polarizers and photon detectors. By rotating the polarizer up to 90 degrees, players can increase their winning probabilities significantly compared to classical methods, enhancing average win rates with quantum strategies. Looking at Fig.~\ref{FIG2b} and depending on the discussion above, it seems counter-intuitive that initially when the number of nodes is below 2, classical games perform better than the quantum version. Its worth-noting here that quantum games need at least two nodes in the network to begin with to distribute an entangled pair of initial quantum states. Therefore, the quantum advantage is visible once the number of nodes in the network increases to more than 2. {To summarize, classical coin flipping relies on a straightforward process where one party flips the coin, and the other predicts the outcome. In quantum coin flipping, the introduction of superposition adds a unique dimension. In classical systems, a coin is definitively in a heads or tails state, but in quantum systems, it exists in a superposition of both states until measured. This quantum uncertainty, absent in classical systems, is crux of quantum coin flipping's advantage. The inherent unpredictability introduced by superposition in quantum systems offers additional degrees of freedom for achieving randomness compared to classical coin flipping.} 

In Fig.~\ref{FIG4}, we analyze performance in terms of the end-to-end fidelity over the network topology in Fig.~\ref{FIG3} as a function of average link decoherence rate. Decoherence in quantum links is represented in terms of damping (amplitude or phase) or depolarizing. The results generated in Fig.~\ref{FIG4} consider depolarizing rate over the quantum links. Quantum strategies perform better owing to the increase in the average equilibrium payoff offered by quantum correlation and unitary strategies. {Quantum advantage in quantum games, as evident in Figs.~\ref{FIG2a}, \ref{FIG2b} and \ref{FIG4} arises from the distinctive principles and phenomena of quantum mechanics, offering players outcomes that cannot be achieved or replicated using classical game theory. Quantum superposition allows players to explore multiple strategies simultaneously, leveraging the ability of a quantum state to exist in a superposition of different and multiple qubit states. Entanglement correlates states of distant particles, enhancing coordination in joint quantum strategies. Quantum measurements introduce unpredictability, complicating predictions for classical opponents.}

\section*{Concluding Remarks and Future Directions}

In this paper, we explored the promise of quantum game-theoretic framework for distributing entanglement within quantum networks, with the aim of striking a flexible balance between link fidelity and latency while maintaining entanglement rate over link coherence. In future, we will generalize the quantum game framework to address different challenges in designing, developing and deploying quantum networks. An interesting direction will be to address the impact of (complex) network topologies on the evolution of the strategy and utility function of a game, or vice-versa studying how different game strategies impact the network topology. 

\par Another promising research thread would be focused on evolutionary game theory \cite{7} which embodies a suitable framework for analyzing co-evolution, i.e., the process in which the properties of interacting systems evolve in dependence of each other at the backdrop of the dynamic fitness landscape. The concepts of spatial structure and evolutionary game theory can be used to understand cooperation and competition among the nodes of a quantum network, regarding the use of quantum resources such as entanglement and the study of the co-evolution of the quantum nodes in response to their environment.
\par Quantum decoherence is a crucial quantum phenomenon that we need to consider when designing optimal resource sharing and allocation algorithms for probabilistic quantum networks. Important open questions in this context relate to how to decide when and where to send information over the network while having only a local $k$-hop knowledge of the network topology (or a limited network state knowledge in general), for example how we should make use of quantum repeaters or how to do error correction while in storage to keep logical qubits alive. Factoring in such aspects related to decoherence in a quantum game theory framework is another interesting line of attack for future work.

\end{document}